\documentclass[11pt]{article}

\makeatletter
\@addtoreset{equation}{section}
\makeatother

\usepackage{epsfig,amsfonts}
\usepackage[fleqn]{amsmath}
\usepackage{amsthm,amssymb}
\usepackage{mathrsfs}
\usepackage{graphicx}
\usepackage{hhline}
\usepackage{cite}
\usepackage[dvips]{color}
\usepackage{upgreek}
\newcommand{\betab}{\mbox{\boldmath$\beta$}}
\newcommand{\gammab}{\mbox{\boldmath$\gamma$}}

\def\be{\begin{equation}}
\def\ee{\end{equation}}
\def\bea{\begin{eqnarray}}
\def\eea{\end{eqnarray}}

\def\nfrac#1#2{\genfrac{}{}{0pt}{}{#1}{#2}}

\def\XXint#1#2#3{{\setbox0=\hbox{$#1{#2#3}{\int}$}
    \vcenter{\hbox{$#2#3$}}\kern-.5\wd0}}

\def\zb{\bar z}

\def\pz{\partial_z}
\def\pzb{\partial_{\bar z}}

\def\s{\sigma}

\def\Tb{\bar T}
\def\Thb{\bar \Theta}

\def\tx{\text{x}}
\def\ty{\text{y}}
\def\tz{\text{z}}
\def\tzb{\bar{\text{z}}}

\def\tF{\text F}

\def\cO{\mathcal{O}}
\def\cA{\mathcal{A}}

\begin{document}
%\draft

%%%%%%%%%%%%%%%%%%% Title %%%%%%%%%%%%%%%%%%%%%%%%%%%%%%%%%%%%%%%%%%%

\begin{titlepage}

\vspace{0.0cm}

\begin{center}
\begin{LARGE}
{\bf On space of integrable quantum field theories}

\vspace{0.5cm}

\end{LARGE}

\vspace{0.8cm}

\begin{large}

{\bf F.A. Smirnov$^{1}$, A.B. Zamolodchikov}$^{2,3}$

\end{large}

\vspace{0.8cm}

{{
${}^{1}$ Sorbonne Universit\'e, UPMC Univ Paris 06\\ CNRS, UMR 7589, LPTHE\\F-75005, Paris, France}\\

${}^{2}$ NHETC, Department of Physics and Astronomy\\
     Rutgers University\\
     Piscataway, NJ 08855-0849, USA\\

\vspace{.2cm}

${}^{3}$ Institute for Information Transmission Problems\\
127051 Moscow, Russia}

\vspace{0.8cm}

\centerline{\bf Abstract} \vspace{.8cm}
\parbox{14.0cm}{

We study deformations of 2D Integrable Quantum Field Theories (IQFT) which preserve integrability (the existence of infinitely many local integrals of motion). The IQFT are understood as "effective field theories", with finite ultraviolet cutoff. We show that for any such IQFT there are infinitely many integrable deformations generated by scalar local fields $X_s$, which are in one-to-one correspondence with the local integrals of motion; moreover, the scalars
$X_s$ are built from the components of the associated conserved currents in a universal way.
The first of these scalars, $X_1$, coincides with the composite field $(T{\bar T})$ built
from the components of the energy-momentum tensor. The deformations of quantum field theories generated by $X_1$ are "solvable" in a certain sense, even if the original theory is not integrable. In a massive IQFT the deformations $X_s$ are identified with the deformations
of the corresponding factorizable S-matrix via the CDD factor. The situation is illustrated
by explicit construction of the form factors of the operators $X_s$ in sine-Gordon theory.
We also make some remarks on the problem of UV completeness of such integrable deformations.

\smallskip

This paper is an extended version of the talk given at the Simons Center, 2015-03-04,  {http://media.scgp.stonybrook.edu/presentations/20150304\_{\rm Zamolodchikov.pdf}}
}
\end{center}

\begin{flushleft}
\rule{4.1 in}{.007 in}\\
{August 2016}
\end{flushleft}
\vfill

\end{titlepage}
\newpage

\section{Introduction}

A substantial number of Integrable Quantum Field Theories (IQFT) is known in two space-time dimensions. If $\Sigma$ is the space of all 2D Quantum Field Theories (QFT), one can think of the subspace $\Sigma^\text{Int} \subset \Sigma$ of IQFT. This paper is an attempt to get insight into the geometry of $\Sigma^\text{Int}$. Given an IQFT, we will try to enumerate all its
infinitesimal deformations which preserve integrability. By definition, such deformations form the tangent space $T\Sigma^\text{Int}|_\text{IQFT}$, which is a subspace of  $T\Sigma|_\text{IQFT}$. We mostly ignore the profound question of ultraviolet (UV) completeness, assuming that the theory has UV cutoff. One can think of elements of $\Sigma$ as "effective
field theories" which make sense only at sufficiently large length scales. Then the space $T\Sigma|_\text{IQFT}$ is given by the span of all local scalar fields (modulo total derivatives) present in a given IQFT, and the subspace $T\Sigma^\text{Int}|_\text{IQFT}$ consists of all fields which, being added as perturbations of IFT, preserve its integrability.

In this work we show that for any IQFT the space $T\Sigma^\text{Int}|_\text{IQFT}$ includes an infinite number of independent fields $X_s$, where $s$ runs the values that enumerate the commuting local integrals of motion (IM)  of the IQFT. The scalars $X_s$ can be defined in terms of the components of the local currents associated with the corresponding IM (Sect.4). Alternatively, in massive theories, the fields $X_s$ are identified with special solutions of the form factor bootstrap equations, relating them to the deformations of the factorizable S-matrix via the CDD factor (Sect.7). In many cases the set $\{X_s\}$ form basis in $T\Sigma^\text{Int}|_\text{IQFT}$, but generally a finite number of additional fields have to be added to span the whole of this space. We illustrate the situation by explicit construction in the case of the sine-Gordon model (Sect.8).

The field $X_1$ is identical to the composite field $(T{\bar T})$. We show that the deformations generated by $X_1$ are "solvable" in a certain sense, even if the original theory is not integrable, and we discuss some properties of such deformations (Sect.5).

Our calculations in Sect.4 and 6 suggest that the question of integrability can be untangled
from the problem of UV completeness. Our statements below apply to QFT understood as  "effective field theories", in which all UV pathologies can be hidden under a short-distance cutoff. However, in Sect.9 we make some remarks concerning possible UV behavior of IQFT, and on the problem of UV completeness.

\section{QFT and deformations}

In this discussion, QFT is understood in an abstract sense, as an
infinite vector space $\mathcal{F} = Span\{\cO_a(z)\}$ of local fields, and
a collection of their correlation functions
\begin{eqnarray}\label{cf0}
\langle\,\cO_{a_1}(z_1)\ ... \ \cO_{a_n}(z_n) \,\rangle\,.
\end{eqnarray}
Here $z$ (e.g. $z_1, ...,z_n$ above) generally denotes point of
space-time. Anticipating dealing with IQFT, we limit our attention to
2D space-time, which we take to be Euclidean. Then the points $z$ can be labeled by complex coordinates, which are denoted $\tz, \tzb$,
\begin{eqnarray}
z \ \to\ (\tz, \tzb)\,, \qquad \bigg\{ \nfrac{\tz=\tx+i\ty}{\tzb =\tx-i\ty}
\end{eqnarray}

The correlation functions \eqref{cf0} are required to satisfy certain properties (some of which we specify below, as needed), the most important being the Operator Product Expansion (OPE)\footnote{Here the OPE is understood in the strong sense: we assume that \eqref{ope} converges at some finite range of separations $z_1-z_2$. Then the bi-local products as in the l.h.s. of \eqref{ope} can be understood as elements of $\mathcal{F}$. Although this assumption is not crucial for our conclusions below, it considerably shortens some of our arguments.}
\begin{eqnarray}\label{ope}
\cO_a(z_1) \cO_b(z_2) = \sum_{c}\,C_{ab}^c(z_1-z_2)\,\cO_c(z_2)\,,
\end{eqnarray}

One might (or might not) think in terms of Lagrangian QFT, where the theory is  described by some sort of local action
\begin{eqnarray}\label{action1}
\cA[\varphi] = \int\,\mathcal{L}\left(\varphi(z), \partial_\mu \varphi(z), \partial_\mu\partial_\nu\varphi(z), ...\right)\,d^2 z
\end{eqnarray}
which appears in the functional integral over a set of "fundamental fields" $\varphi$.
As was mentioned, in this discussion we mostly ignore the problem of UV completeness of the theory, assuming that some UV regularization (with the microscopic cutoff distance $\epsilon$) is imposed. The density $\mathcal{L}$ may involve higher derivatives of the fundamental fields, i.e. \eqref{action1} is a generic {\it quasi-local} action in the sense of Ref.\cite{wilson}. In this approach the space $\Sigma$ is understood as the space of quasi-local actions \eqref{action1}. The coordinates $\{g^i\}$ on $\Sigma$ may be given by a full set of parameters - "coupling constants" - on which $\mathcal{L}(\varphi(z), ... \,| g^i)$ may depend. To shorten notations, we denote by $\mathcal{A}_g$, $g = \{g^i\}$, the points of $\Sigma$. Let again $\mathcal{F}_g$ be the space of local fields in $\mathcal{A}_g$. Generic variation of the action \eqref{action1} can be written as
\begin{eqnarray}\label{varaction1}
\delta\cA = \int\,\delta\mathcal{L}(z)\,d^2 z \,, \qquad \delta\mathcal{L}(z) = \sum_i\,\delta g^i\,O_i(z)\,,
\end{eqnarray}
where $O_i(z)$ are elements of a basis in the factor-space
\begin{eqnarray}
{\hat{\mathcal{F}}}_g = \mathcal{F}^{(0)}_g/\partial\mathcal{F}_g\,;
\end{eqnarray}
with $\mathcal{F}^{(0)}_g$ being the subspace of {\it scalar} fields (for simplicity, we assume that $\Sigma$ includes only rotationally invariant theories), and $\partial\mathcal{F} = Span\{\partial_\tz \cO_a, \partial_{\bar\tz} \cO_a\}$ is the subspace of total derivatives, which bring zero contributions to to the integral in \eqref{varaction1}.
The Lagrangian approach formulation makes self evident the following {\it deformation formula}
\begin{eqnarray}\nonumber
\delta_g \langle\,\cO_{1}(z_1) \cdots \cO_{n}(z_n)\,\rangle_g =& -& \sum_{i} \delta g^i
\int\,d^2 z \langle\,O_i(z)\cO_{1}(z_1) \cdots \cO_{n}(z_n)\,\rangle_g\\
\label{defformula}&+ &\sum_{k=1}^n
\langle\,\cO_{i_1}(z_1) \cdots\delta_g\cO_{i_k}(z_k) \cdots \cO_{i_n}(z_n)\,\rangle_g
\end{eqnarray}
Here $\delta_g \cO_a = \sum_i \delta g^i \,({\hat B}_{i}(g) \cO)_a$, where
$B_i(g)$ are some linear operators in $\mathcal{F}_g$. (The integral
over $z$ can - and usually does - diverge as $z\to z_k$, and in UV complete theory $\delta \cO_k(z_k)$ must include cutoff dependent counterterms to make the  finite limit $\epsilon\to 0$ possible.) In what follows we will not explicitly refer to any Lagrangian representation, but simply postulate the above deformation formula.
The latter then represents the sense in which the space ${\hat{\mathcal{F}}}_g$ is the tangent one $T\Sigma\big|_g$.

\section{IQFT and local IM}

One of the common characteristics of Integrable Field Theories is the presence of an infinite set of commutative local Integrals of Motion (IM).
Local IM are generated by local currents, i.e. pairs of local fields $(T_{s+1}(z), \Theta_{s-1}(z))$, which satisfy the continuity equations
\begin{eqnarray}\label{Tcont}
\partial_{\tzb} T_{s+1}(z) = \partial_{\tz} \Theta_{s-1}(z)
\end{eqnarray}
The index $s$ labels the currents; we will assume its values to represent their spins: the subscripts $s+1$ and $s-1$
indicate the spins of the corresponding fields\footnote{This identification is convenient, but not essential for our arguments below. When there are more than one current of the same spin, additional labels may be introduced.}. The spins $s$ for the currents takes values in some set $\{s\} \subset \mathbb{Z}$ which may be different for different IQFT\footnote{Since we assume that the currents are local fields, only integer or half-integer spins $s$ are allowed, but the requirement of commutativity \eqref{Pcomm} rules out the possibility of having many
fermionic elements. Supersymmetry provides an interesting extension, but we do not discuss it here.}. However, in all QFT there are conserved currents \eqref{Tcont} with $s=\pm 1$, the components of its energy-momentum tensor $T_{\mu\nu}$. Below we also use conventional notations
\begin{eqnarray}\label{emtensor}
T = -2\pi\ T_{\tz\tz}\,, \quad \Tb =-2\pi\ T_{\tzb \tzb}\,,
\quad\Theta = 2\pi\ T_{\tz \zb}
\end{eqnarray}
for these components. If the theory is P-invariant (which we assume), the set $\{s\}$ is symmetric with respect to the P-reflection $s\leftrightarrow -s$. In what follows it will be convenient to use separate notations for negative $s$, i.e. for $s>0$ we write $\Theta_{-s-1}$ as $\Tb_{s+1}$, and $T_{-s+1}$ as  $\Thb_{s-1}$, and remove all negative entries from $\{s\}$. The negative-$s$ equations \eqref{Tcont}
then read
\begin{eqnarray}\label{Tbcont}
\partial_\tz \Tb_{s+1}(z) = \partial_{\tzb}\Thb_{s-1}(z)\,.
\end{eqnarray}
It follows from \eqref{Tcont} and \eqref{Tbcont} that the integrals
\begin{eqnarray}\label{Psdef}
&&P_s = \frac{1}{2\pi}\,\int_C \,T_{s+1}(z)\,d\tz + \Theta_{s-1}(z)\,d\tzb\\
&&{\bar P}_s = \frac{1}{2\pi}\,\int_C \,\Tb_{s+1}(z)\,d\tzb + \Thb_{s-1}(z)\,d\tz\label{Pbsdef}
\end{eqnarray}
do not change under trivial deformations of the integration path $C$, and thus define local IM.

The notion of integrability requires that the operators\footnote{As usual, the space of states and operator representation may depend on the choice of the Hamiltonian picture (equal-time slices); specifics of such choice are completely irrelevant for the present discussion.}
$P_s$ form a commutative set,
\begin{eqnarray}\label{Pcomm}
[P_s, P_{s'}] =0
\end{eqnarray} for any $s,s' \in \{s\}$.
For local IM of the form \eqref{Psdef}, \eqref{Pbsdef} this condition implies
\begin{eqnarray}\label{PTcomm1}
[P_\s, T_{s+1}(z)] = \pz A_{\s,s}(z)\,,
\qquad [P_\s, \Theta_{s-1}(z)] = \pzb A_{\s,s}(z)\,,
\end{eqnarray}
and
\begin{eqnarray}\label{PTcomm1b}
[P_\s, {\Tb}_{s+1}(z)]  = \pzb B_{\s,s}(z)\,, \qquad
[P_\s, {\Thb}_{s-1}(z)]  = \pz B_{\s,s}(z)\,,
\end{eqnarray}
where $A_{\s,s}$ and $B_{\s,s}$ are some local fields, as well as similar equations for the commutators of ${\bar P}_s$ with local currents.
Let us remind here that the commutators $[P_s, \cO(z)]$ with any local field $\cO$ can be defined, in the Euclidean language, by the integrals
\begin{eqnarray}\label{commutator}
[P_s,O(z_0)] = \frac{1}{2\pi}\,\oint_{C_{z_0}}[T_{s+1}(z)d\tz+\Theta_{s-1}(z)d\tzb] \,O(z_0)
\end{eqnarray}

In QFT with "coventional" UV behavior (i.e. the one controlled by some UV fixed point) the components $(T_{s+1}, \Theta_{s-1})$ both have scale dimensions $s+1$. Let us note here that generally the dimensions are defined relative to to a given fixed point; Generally, if QFT has a more complicated UV structure, the notion of dimensions may be ambiguous.

\section{Local fields $X_s$}

Given the currents $(T_{s+1},\Theta_{s-1})$ and $(\Tb_{s+1},\Thb_{s-1})$, one may attempt to construct "composite" scalar fields
by taking limit $z\to z'$ in the operator products $T_{s+1}(z)\Tb_{s+1}(z')$ and $\Theta_{s-1}(z)\Thb_{s-1}(z')$. Of course generally
such limits are singular, demanding subtractions and thus making the result ambiguous. It turns out that if one takes OPE in the special
combination
\begin{eqnarray}\label{tt-thth}
T_{s+1}(z)\Tb_{s-1}(z')-\Theta_{s-1}(z)\Thb_{s-1}(z')
\end{eqnarray}
 and ignores the terms with total derivatives of local fields in the
expansion, the non-derivative divergent terms cancel out (in fact, all non-derivative terms with singular coefficients disappear in the OPE of
\eqref{tt-thth}, see below). As the result, the limit $z'\to z$ exists in a straightforward sense, and it uniquely (up to derivatives)
defines the scalar local field $X_s$,
\begin{eqnarray}\label{TTblimit}
\lim_{z\to z'}\left(T_{s+1}(z)\Tb_{s-1}(z')-\Theta_{s-1}(z)\Thb_{s-1}(z')\right) = X_s(z')  \ + \  \text{derivatives terms.}
\end{eqnarray}
Let us note that the "derivative terms" may well involve divergent coefficients, therefore this definition of $X_s$ is unambiguous only
up to the derivatives. Fortunately we are interested in the fields as the vectors in ${\hat{\cal F}}$, so the derivatives are irrelevant.

Let us show that the limit in \eqref{TTblimit} indeed exists. The following calculations are nearly identical to those presented in
Ref.\cite{Z2004}. Consider, say, the $\tzb$ derivative of the combination \eqref{tt-thth}. As the consequence of the
continuity equations \eqref{Tcont}, \eqref{Tbcont}, the following easily verified identity holds
\begin{eqnarray}\label{zbidentity}
&&\partial_{\tzb}\left(T_{s+1}(z)\Tb_{s+1}(z') - \Theta_{s-1}(z)\Thb_{s-1}(z')\right) =\nonumber\\
&&\qquad\qquad\left(\partial_\tz+\partial_{\tz'}\right)\Theta_{s-1}(z)\Tb_{s+1}(z') - \left(\partial_{\tzb} + \partial_{\tzb'}\right)\Theta_{s-1}(z)\Thb_{s-1}(z')\,.
\end{eqnarray}
Now, plug in the OPE of the products appearing in the r.h.s., e.g.
$\Theta_{s-1}(z)\Tb_{s+1}(z') = \sum_i c^i (z-z')\,O_i(z')$, where the sum is over the complete set of independent fields $O_i$
of the theory, and $c^i (z-z')$ are c-number coefficient functions. Since all the
coefficient functions depend on the separation $z-z'$, and thus get annihilated by the derivatives $\partial_\tz+\partial_{\tz'}$ and
$\partial_{\tzb} + \partial_{\tzb'}$ in \eqref{zbidentity}, one concludes that the OPE of the l.h.s. in \eqref{zbidentity} consists entirely of the derivative terms. Similar calculation reveals that the $\partial_{\tz}$ derivative of \eqref{tt-thth} also involves only derivatives of local fields.
That is, both $\partial_{\tzb}$ and $\partial_\tz$ of the operator product \eqref{tt-thth} vanish as the vectors in ${\hat{\cal F}}=\mathcal{F}/{\partial\mathcal{F}}$. In turn, it follows that the OPE of \eqref{tt-thth} consists mostly of the derivative terms, except for a single term which comes with a constant (independent of $z-z'$) coefficient. The value of the coefficient is irrelevant,
since it can be absorbed in the normalization of the field $X_s$ below. Setting this coefficient to one (for every $s$) brings the OPE of
\eqref{tt-thth} to the form
\begin{eqnarray}\label{Xdef}
T_{s+1}(z)\Tb_{s-1}(z')-\Theta_{s-1}(z)\Thb_{s-1}(z') = X_s(z') + \text{derivative terms}\,,
\end{eqnarray}
which in particular makes obvious the regular nature of the limit in \eqref{TTblimit}. Let us stress that the scalar fields $X_s$ $s\in\{s\}$ can be constructed in any IQFT explicitly, in terms of the local IM densities.

Some useful properties of these operators are worth noting. Consider IQFT in the geometry of an infinite cylinder, with the spatial coordinate $\tx$ compactified on a circle with some finite circumference $R$ (Fig.1). Then the
energy spectrum is discrete, and the stationary states $\mid n \,\rangle$
are, generally, non-degenerate. Then, by repeating the arguments in \cite{Z2004},
one can prove that
\begin{eqnarray}\label{nXn}
\langle n\mid X_s\mid n \rangle = \langle n\mid T_{s+1}\mid n \rangle\,
\langle n\mid \Tb_{s+1}\mid n \rangle - \langle n\mid \Theta_{s-1}\mid n \rangle\,
\langle n\mid \Thb_{s-1}\mid n \rangle\,.
\end{eqnarray}
It follows, in particular, that in the infinite system $R=\infty$ expectation values $\langle X_s\rangle$ with $s>1$ vanish, because
the rotational symmetry of the infinite system forces the expectation values in the r.h.s. of \eqref{nXn} to vanish.

In IQFT with UV limit controlled by a CFT, the scalars $X_s$ have the dimensions of $[\text{mass}]^{2(s+1)}$, in particular, they all are "irrelevant" in standard nomenclature.
This simply means that adding such fields as the perturbations to the action (as we do in Sect.6 below) alters the UV properties of the theory, and generally - but not always - breaks UV completeness of the theory. We make further remarks on this point in Sect.9.

\section{${\bf (T{\bar T})}$ flow}

The operator $X_1$ is special. It is built from the components \eqref{emtensor} of the conserved energy-momentum tensor, which is present in any QFT, integrable or not. It is identical to the field $(T{\bar T})$, a frequent actor in a number of previous studies \cite{AlZmassless,x, AlZ-TTbar3, Z2004, tateo1, delfino}. A general definition can be found in \cite{Z2004}. Here we use the notations $X_1$ and $(T{\bar T})$ interchangeably. Infinitesimal deformations generated by this operator turn out to be in some sense "solvable", even if the original theory is not integrable. Consider a curve $\mathcal{A}_\alpha$ in the theory space $\Sigma$, with $\alpha$ denoting the parameter along the curve, such that at any point of the curve the tangent vector is proportional to $X_1$,
\begin{eqnarray}\label{alphaflow}
\frac{d}{d\alpha}\mathcal{A}_\alpha = \frac{1}{\pi^2}\,\int\,(T{\bar T})_\alpha\,d^2 z
\end{eqnarray}
where the additional subscript $\alpha$ in the r.h.s. is added to emphasize
that the operator is built (according to \eqref{Xdef}) from the components of the
energy-momentum tensor associated with the QFT $\mathcal{A}_\alpha$, and the numerical coefficient $1/\pi^2$ is introduced for future convenience. Consider the theory $\mathcal{A}_\alpha$ in finite size geometry, as in Fig.1, and let $E_n=E_n(R,\alpha)$ energies of the stationary states $\mid n\,\rangle$; we also denote $P_n=P_n(R) = 2\pi l_n/R$, $l_n\in \mathbb{Z}$, the corresponding spatial momenta of these states. Then, as was shown in \cite{Z2004}, the equation \eqref{nXn} with $s=1$ reduces to
\begin{eqnarray}
\langle n\mid (T{\bar T}) \mid n\rangle = -\frac{\pi^2}{R}\,\left(E_n \frac{\partial}{\partial R} E_n + \frac{P_n^2}{R} \right)\,.
\end{eqnarray}
Since, by definition of $\mathcal{A}_\alpha$, an infinitesimal shift of $\alpha$ is generated by $(T{\bar T})_\alpha$, this leads to closed differential equation for
the energy levels,
\begin{eqnarray}\label{burgers}
\frac{\partial}{\partial \alpha}E(R,\alpha) + E(R,\alpha)\,\frac{\partial}{\partial R} E(R,\alpha) +\frac{P^2(R)}{R}=0\,.
\end{eqnarray}
The equation has the same form for all eigenvalues $E_n(R,\alpha)$; for this reason we dropped the index $n$. Since in what follows $\alpha$ is generally regarded as the parameter, we use instead the abbreviated notation
\begin{eqnarray}
E_\alpha(R) \equiv E(R,\alpha)
\end{eqnarray}
for any level $E_n(R,\alpha)$.  Let us note here some properties of the solutions of equation \eqref{burgers}, comparing them to the expected properties of the finite-size energies.

(1) Equation \eqref{burgers} has the form of the inviscid Burgers equation with the additional driving force $-P^2(R)/R=-(2\pi l)^2/R^3$, with $E(R,\alpha)$ playing the role of the velocity field, and $\alpha$ interpreted as the time.

(2) The general solution at $P=0$ is well known; it is given explicitly by the equation
\begin{eqnarray}\label{Ealpha}
E_{\alpha}(R) = E_{0}\left(R-\alpha\,E_{\alpha}(R)\right)\,.
\end{eqnarray}
At generic $P=2\pi l/R$ the solution is more complicated but still can be found by the method of characteristics.

(3) Since in our context the solution $E_{\alpha}(R)$ has the meaning of the finite size energy levels, one expects them to behave as $E_{\alpha}(R) \simeq F_{\alpha}\,R$, up to the terms bounded at $R\to\infty$. The above equation yields the $\alpha$-dependence of the bulk vacuum energy density $F_\alpha$
\begin{eqnarray}
F_{\alpha} =\frac{F_0}{1+\alpha F_0}\,,
\end{eqnarray}
where $F_0$ is the vacuum energy density of the unperturbed theory $\mathcal{A}_0$. Furthermore, if the theory is massive, it follows from \eqref{burgers} that the mass $M_\alpha$ of any of its particles depends on $\alpha$ as\footnote{It is easy to check that $E_\alpha=F_\alpha R + \sqrt{M_\alpha^2 + P^2(R)}$ solves \eqref{burgers}, provided $M_\alpha$ is given by \eqref{Malpha}.}
\begin{eqnarray}\label{Malpha}
M_\alpha = \frac{M_0}{1+\alpha\,F_0}\,.
\end{eqnarray}
Since this equation applies to any particle of the theory, the mass ratios in $\mathcal{A}_\alpha$ are independent of $\alpha$.

(4) In general, the energy levels can be written as
\begin{eqnarray}
E_\alpha = F_\alpha R + \mu_\alpha\,u(r,t)\,,
\end{eqnarray}
where $r=\mu_\alpha R$, $\mu_\alpha$ is an $\alpha$-dependent mass scale which satisfies $\mu_\alpha= \mu_0/(1+\alpha F_0)$ (in massive theories one can take $\mu_\alpha=M_\alpha$), $t=\alpha\,\mu_0 \mu_\alpha$, and the dimensionless function $u(r,t)$ is bounded as $r\to \infty$. Then it is straightforward to check that $u=u(r,t)$ itself satisfies the same equation
\begin{eqnarray}\label{burgers1}
\partial_t u + u\,\partial_r u+\frac{(2\pi l)^2}{r^3}=0\,,
\end{eqnarray}
in terms of the dimensionless quantities.

(5) If the theory is massive, one can consider finite size energy levels corresponding to
two identical particles having opposite momenta $p$ and $-p$ (so that the total momentum $P$ is zero). If $M R >> 1$ ($M=M_\alpha$),
and energies are well below all inelastic thresholds, the $R$-dependence of of $E=E(R,\alpha)$
has the form
\begin{eqnarray}\label{twopart}
E = F_\alpha\,R + 2\,\sqrt{M^2+p^2}\,,
\end{eqnarray}
up to terms exponentially small in $R$. The momentum $p$ here is subject to the quantization condition $p R + \Delta(p)= 2\pi n$,
where $\Delta(p)=\Delta_\alpha(p)$ is the scattering phase. It is not difficult to show that consistency of \eqref{twopart} with \eqref{burgers1} demands $\Delta_\alpha (p) = \Delta_0(p) - 2\alpha\,p\,\sqrt{M^2+p^2}$, or, in terms of the rapidity difference $\theta=\theta_1-\theta_2$
($\theta$ parameterizes the particle momentums as $p=M\,\sinh(\theta/2)$)
\begin{eqnarray}\label{Delta-alpha}
\Delta_\alpha = \Delta_0 - \alpha M^2\,\sinh\theta\,.
\end{eqnarray}
We see that the effect of the $\alpha$-flow on the two particle S-matrix is in adding a CDD factor
$S(\theta)=S_0(\theta)\,\exp\{-i\alpha M^2\,\sinh\theta\}$ (this of course agrees with Eq.\eqref{cdd-deformations} in sect.7 below, provided one identifies $\alpha M^2 = -\alpha_1$).

If the theory $\mathcal{A}_0$ is integrable, the theories $\mathcal{A}_\alpha$ are integrable as well (see sect.6 below). In that case the ground state energy of the finite-size system can be found using the Thermodynamic Bethe Ansatz (TBA) technique \cite{AlZ}(for a review see e.g \cite{MussardoPhysRep}). It is not difficult to show that
the deformation \eqref{Delta-alpha} of the scattering phase in the TBA equations leads to the deformation of the ground state energy according to \eqref{Ealpha}.

(6) As is well known, solutions of Eq.\eqref{burgers} tend to develop "shocks". Mathematically, shocks are algebraic (square-root) singularities of $E(R,\alpha)$. Even if one starts with physically acceptable $E(R,0)$ (the analytic function of $R$ at all real $R>0$, with singularity
at $R=0$), evolution in $\alpha$ may generate a shock singularity at finite positive $R$. A simple example is provided by the case when $\mathcal{A}_0$ is CFT, where the finite-size energies of the stationary states $\mid n\rangle$ have the standard form
\begin{eqnarray}
E_0(R) = F_0\,R - \frac{C}{R}\,,
\end{eqnarray}
 with the constants $C=C_n = (\pi/6R)\,(c-12(\Delta_n+{\bar\Delta}_n))$ expressed in terms of the central charge $c$ and the eigenvalues $\Delta, {\bar\Delta}$ of the operators $L_0, {\bar L}_0$. Limiting our attention to the case $P_n=0$ (i.e. $\Delta={\bar\Delta}$) one finds from
 \eqref{Ealpha}
 \begin{eqnarray}\label{CFTflow}
 E_\alpha(R) = F_\alpha\,R + \frac{R}{2\tilde\alpha}\,\left(1-\sqrt{1+\frac{4{\tilde \alpha}\,C}{R^2}}\right)\,,
 \end{eqnarray}
where ${\tilde\alpha}=\alpha\,(1+\alpha\,F_0)$. When ${\tilde\alpha}\,C_n$ is negative,
$E_n(R,\alpha)$ develops a square-root singularity at real positive $R=2\sqrt{-{\tilde\alpha} C_n}$.
On the other hand, if ${\tilde \alpha} C$ is positive, $E_\alpha(R)$ is free from singularities
at all real $R^2$, including $R^2=0$. In fact, it is easy to argue that these features of $E_\alpha(R)$ are not specific to the cases when $\mathcal{A}_0$ is a CFT. This follows from an alternative form of Eq.\eqref{Ealpha},
\begin{eqnarray}
R_\alpha = R_0 + \alpha\,E
\end{eqnarray}
in terms of the functions $R_0(E),\  R_\alpha(E)$ inverse to $E_0(R),\ E_\alpha(R)$, respectively  (we still assume $P=0$ for simplicity, and regard $\alpha$ as a parameter). It shows that the $E$ vs $R$ plots of $E_\alpha(R)$ and $E_0(R)$ are related just by affine transformation of the coordinate axes $E\to E, \ R\to R-\alpha E$, hence the above features are typical if one assumes that $E_0(R)$ is regular at all $R>0$ but diverges at $R=0$. Both at positive and negative ${\tilde\alpha} C$, the
form \eqref{CFTflow} looks pathological, or at least unusual, if one wants to interpret $E_\alpha(R)$ in terms of local QFT with finitely many local degrees of freedom. But while the $R^2\to 0$ behavior at positive ${\tilde \alpha} C$ may, in principle, be excused
in a theory with finite UV cutoff\footnote{In fact, it is possible to argue that the theory $\mathcal{A}_\alpha$ with positive $\alpha$, even if equipped with finite UV cutoff, does not have a
ground state at any $R$, at least if $c$ is positive. We will comment on this point elsewhere \cite{SZ2}.}, the singularity at finite positive $R$ is more troublesome. A possible connection of such "shock" singularities to the problem of UV completeness of the theory $\mathcal{A}_\alpha$, Eq.\eqref{alphaflow}, will be discussed in Sect.9 in a more general context.
Here we note that the finite size spectrum \eqref{CFTflow} was obtained in \cite{dub1,dub2,tateo1},
as the solution of Nambu string quantized in a certain unitary gauge, under conjecture that
interactions of transverse string oscillations are described by $(T{\bar T})$ perturbation of free bosonic CFT, and using techniques of IQFT with the two-particle S-matrix $S(\theta)=\exp\{-i\alpha M^2\,\sinh\theta\}$.

The results presented in this section have substantial overlap with interesting recent
work of Cavagli\'a, Negro, Szecsenyi, and Tateo \cite{tateo2}.

\section{Integrable Perturbations}

We now want to show that every field $X_s \in{\hat{\mathcal{F}}}$ generates an integrable deformation of a given IQFT, or, in other words, that $X_s$ all
lie in $T\Sigma^\text{Int}\big|_\text{IQFT}$, the tangent to the subspace
$\Sigma^\text{Int} \subset \Sigma$ at the IQFT.

To this end, let us first prove that the commutator of any local IM $P_\sigma$ with any of the fields $X_s(z)$ is a total derivative of a local field, i.e.
\begin{eqnarray}\label{PXcomm}
[P_\sigma, X_s(z)] \ \in \ \partial\mathcal{F}\,.
\end{eqnarray}
Here the commutator can be understood as the contour integral \eqref{commutator}. To prove \eqref{PXcomm}, replace $X_s(z)$ by its point-splitted version, and consider the commutator
\begin{eqnarray}\label{Pcommsplit}
\left[P_\sigma, \left(T_{s+1}(z)\Tb_{s+1}(z') - \Theta_{s-1}(z)\Thb_{s-1}(z')\right)\right]\,.
\end{eqnarray}
The commutator naturally splits into "$z$-terms", the terms generated by commuting $P_\sigma$ with the densities localized at $z$, and "$z'$-terms"
from the commutations with the densities at $z'$. Recalling the relations
\eqref{PTcomm1} one finds
\begin{eqnarray}
&&\text{"$z$-term"} = \pz A_{\s,s}(z)\,\Tb_{s+1}(z') - \partial_{\bar\tz}A_{\s,s}(z)
\Thb_{s-1}(z') =\\
&&=\left(\pz+\partial_{\tz'}\right)A_{\s,s}(z)\,\Tb_{s+1}(z')-
\left(\partial_{\bar\tz}+\partial_{\bar\tz'}\right)A_{\s,s}(z)
\Thb_{s-1}(z') \ \in \ \partial\mathcal{F}\,,\nonumber
\end{eqnarray}
Similar calculation using \eqref{PTcomm1b} shows that $z'$-term lies in
$\partial\mathcal{F}$ as well.  Therefore, the whole commutator \eqref{Pcommsplit} reduces to a combination of total derivatives,
and the desired result \eqref{PXcomm} follows in the limit $z'\to z$.

Take a generic IQFT (which we denote $\mathcal{A}_0 \in \Sigma^\text{Int}$), and focus on one of its local IM, say $P_\sigma$. Consider the correlation function
\begin{eqnarray}\label{intcorr}
\langle\,\cO\ \oint_C \left[T_{\s+1}(z)d\tz +\Theta_{\s-1}(z) d\tzb\right]\,\rangle
\end{eqnarray}
where $\cO$ stands for any insertion of the form $\cO_{a_1}(z_1) \cO_{a_2}(z_2) ... \cO_{a_n}(z_n)$, and the $\tz, \tzb$ integration is over some closed contour $C$. The continuity equation for the current $(T_{\s+1}, \Theta_{\s-1})$ is equivalent to the statement that \eqref{intcorr} vanishes as long as all the insertion points $z_1, ..., z_n$ lie outside the integration contour $C$.

Now, let $\mathcal{A}_0+\delta_{g_s}\mathcal{A}$ be an infinitesimally close QFT generated by
adding $\delta g_s \,\int X_s(z)\,d^2 z$ to the action. According to
the deformation formula \eqref{defformula}, the associated deformation of
\eqref{intcorr} has the form
\begin{eqnarray}\nonumber
&&\delta_{g_s} \left(\text{Eq}.\eqref{intcorr}\right) = -\,\delta g_s\,\int\,d^2 w\,\langle\,X_s(w)\,\cO\,\oint_C \left[T_{\s+1}(z)d\tz +\Theta_{\s-1}(z) d\tzb\right]\,\rangle +\\
&&\qquad\qquad \langle\,\cO\, \oint_C \left[\delta_{g_s}T_{\s+1}(z)d\tz +\delta_{g_s}\Theta_{\s-1}(z) d\tzb\right]\,\rangle \,, \label{deltaint}
\end{eqnarray}
where we ignored the term with $\delta_{g_s}\cO$ in the r.h.s. since
it plays no role in the analysis below. The deformed theory would still have the IM $P_\s$ if the field variations $\delta_{g_s}T_{\s+1}$ and $\delta_{g_s}\Theta_{\s+1}$ can be adjusted in such a way that the full variation \eqref{deltaint} vanishes.

To see that this is always possible, let us assume for simplicity that the contour $C$ is simple, and focus on the first term in the r.h.s. of \eqref{deltaint}. Split the integration over $w$ into two parts\footnote{Here we ignore possible "contact terms" contributions which may came from the integration region $|w-z|<\epsilon$. It is easy to see that such terms can be absorbed into re-definitions of $\delta_{g_s}T_{\s+1}$ and $\delta_{g_s}\Theta_{\s-1}$ in Eq.\eqref{deltaint}.},
\begin{eqnarray}\label{wsplit}
\int_{\mathbb{R}^2}\,d^2 w \,\,( ... ) = \int_{D(C)}\,d^2 w\,\, ( ... ) +
\int_{{\bar D}(C)}\,d^2 w \,\,( ... ) \,,
\end{eqnarray}
where $D(C)$ is the part of $\mathbb{R}^2$ lying inside $C$, and ${\bar D}$ is the complement of $D$. Then the second term in \eqref{wsplit} vanishes,
because for any fixed $w \in {\bar D}(C)$ the contour $C$ leaves outside all insertion points of $X_s(w)\,\cO$. A nonzero contribution may arise from the first term, where $w$ falls inside $C$. With $w\in D(C)$ fixed, one can collapse the contour $C$ on $w$, thus reducing the integral over $z$ to the commutator $[P_\s, X_s(w)]$ (see \eqref{commutator}), which, according to
\eqref{PXcomm}, lies in $\partial\mathcal{F}$. Thew latter statement means that
\begin{eqnarray}\label{PXcomm1}
4\pi i\,\,[P_\s, X_s(w)] = \partial_{\bar{\text{w}}}{\hat T}_{\s+1, \,s}(w) +
\partial_\text{w}{\hat\Theta}_{\s-1,\,s}(w)\,,
\end{eqnarray}
where ${\hat T}_{\s+1,\,s}$ and ${\hat\Theta}_{\s-1,\,s}$ are some local fields of spins $\s+1$ and $\s-1$, respectively. Thus, in the remaining integral over $w$ the integrand is written as a total derivative, and the integral reduces to the boundary contribution, i.e.
to contour integral over $C$.
As a result, the first term in the r.h.s. of \eqref{deltaint} transforms to
\begin{eqnarray}
-\delta g_s\,\langle\,\cO\,\oint_C \left[{\hat T}_{\s+1,\,s}(z)d\tz +{\hat\Theta}_{\s-1,\,s}(z) d\tzb\right]\,\rangle \,. \label{deltaint1}
\end{eqnarray}
Thus, the full variation \eqref{deltaint} can be made equal to zero by choosing
\begin{eqnarray}
\delta_{g_s}T_{\s+1} = \delta g_s\,{\hat T}_{\s+1, \,s}\,,\qquad
\delta_{g_s}\Theta_{\s-1} = \delta g_s\,{\hat \Theta}_{\s-1, \,s}\,.
\end{eqnarray}
We conclude that after infinitesimal deformation generated by the operator $X_s$ the integral $P_\s$, Eq.\eqref{Psdef} still conserves, provided the densities $(T_{\s+1},\Theta_{\s-1})$ are deformed as
\begin{eqnarray}
&&T_{\s+1} \ \to\ T_{\s+1} + \delta g_s\,{\hat T}_{\s+1,\,s}\,,\\
&&\Theta_{\s-1} \ \to\ \Theta_{\s-1} + \delta g_s\,{\hat \Theta}_{\s-1,\,s}\,.
\end{eqnarray}

Note that the above analysis applies to deformation generated by any of the scalar fields $X_s$, $s\in\{s\}$, or any linear combinations thereof , and demonstrates conservation of the whole set of integrals $P_\s, {\bar P}_\s,\ \s\in\{s\}$ in the deformed theory. However, it does not prove that the deformed IM $\{P_s, {\bar P}_s\}$ still commute with each other. Although at the moment we do not have satisfactory proof of this statement, we
find its general validity very likely. One of the arguments is as follows. Consider two IM, say $P_s$ and $P_{s'}$, and suppose that after the deformation the commutator ceases to be zero, $[P_{s},P_{s'}] = Q_{s+s'} \neq 0$. The operator $Q_{s+s'}$ must be another local IM of the form
\eqref{Psdef}, of the spin $s+s'$. A priori, there are two possibilities. Either $Q_{s+s'}$ is the tdeformation of one of the local IM of the original theory\footnote{This would be impossible for some classes of IQFT. For example, in many cases, such as the sine-Gordon model, the set $\{s\}$ includes only odd integers; in such cases $s+s'$ can not be in $\{s\}$.}, or it is an entirely new IM. In both cases the IM of the deformed theory would form a non-abelian algebra of {\it local} higher spin IM, which would provide extremely powerful symmetry structure, so far unknown outside CFT or free massive QFT.  Therefore, breakdown of the commutativity under deformation
is unlikely: it would be "too good to be true". With this
reasoning, we conjecture that the IM of the deformed theory generally
still commute with each other.

\section{S-matrix and Form-factors}

Typical IQFT is massive\footnote{Exceptions are integrable CFT, and the so called integrable massless flows. The latter correspond to the special (integrable) RG flows ending at IR fixed points. In such cases the notions of particles and $S$-matrix are less physically clear, and generally are  ambiguous. Nonetheless, many such theories admit treatment based by "massless S-matrix" and associated TBA equations \cite{zz3,AlZmassless,MasslessFlows}.}. As any massive theory, it is completely characterized by the associated particle theory - the spectrum of stable particles and $S$-matrix.
The presence of higher-spin local IM forces the S-matrix to be purely elastic, in which the number $N$ of particles and the set of their individual momenta is preserved after the scattering process. Such $S$-matrices are known as "factorizable", because then $N\to N$ $S$-matrix is expressed as the product of $2\to 2$ $S$-matrices.
The latter is the function of a single kinematic variable $\theta=\theta_1-\theta_2$, the difference of the particle's rapidity. In general, in the presence of mass degeneracies
in the particle spectrum, the two-particle $S$-matrix ${\hat S}(\theta)$ is an operator acting in the "flavor" spaces of
the colliding particles (see e.g. \cite{zz1}).

Deformations of IQFT preserving integrability, described in the previous sections, must generate deformations of the factorizable $S$-matrix. To understand the situation, let us recall that in the factorizable scattering theory the two-particle $S$ matrix must
satisfy a number of general conditions. When mass degeneracy is present, ${\hat S}(\theta)$ satisfies the celebrated Yang-Baxter equation, which typically fixes the "flavor" structure up to a finite number of parameters, but leaves the freedom multiplying ${\hat S}(\theta)$ by an arbitrary overall scalar factor. In addition, there are general constraints of analyticity, crossing symmetry, and unitarity, which together fix the scalar factor up to the so-called CDD ambiguity,
\begin{eqnarray}\label{cdd}
{\hat S}(\theta) \ \to \ {\hat S}(\theta)\,\Phi(\theta)
\end{eqnarray}
where $\Phi(\theta)$ is a meromorphic function, which is analytic and bounded in the "physical strip", and satisfies the equations
\begin{eqnarray}\label{cddeqs}
\Phi(\theta)\Phi(-\theta) =1\,, \qquad \Phi(i\pi+\theta)\Phi(i\pi-\theta) = 1\,.
\end{eqnarray}
Thus, a generic CDD factor admits the formal representation
\begin{eqnarray}\label{cddsum1}
\Phi(\theta) = \exp\bigg\{i\,\sum_{s=1}^\infty\,\alpha_s\,\sinh(s\theta)\bigg\}\,.
\end{eqnarray}
In fact, in many cases (like sine-Gordon model, or $O(N)$ sigma models) the crossing symmetry also excludes terms with even $s$ in \eqref{cddsum1}. Although important exceptions exist\footnote{A notable exception is the situation when the particle spectrum contains charge-conjugated pairs of particles $A, {\bar A}$, but all the $A+{\bar A} \to  A+{\bar A}$ scattering amplitudes have a zero "reflection" component (see e.g. \cite{me-z3} for an example). Such structure is compatible with IM $P_s$ with even $s$, since the charge conjugation acts on the local IM as $C P_s C = (-)^{s+1} P_s$. More generally, it is possible that the two-particle $S$-matrix, as an operator in the "flavor" spaces, has block-diagonal structure; in such cases the CDD ambiguity may involve more then one functional factor. In this discussion we ignore such complications.}, here we assume that $\{s\}$ includes only odd entries (equivalently, the CDD factor satisfies $\Phi(\theta)=\Phi(i\pi-\theta)$) to simplify the arguments below. Furthermore, a possible bound-state structure (i.e. identifications of physical poles with particles) may impose additional constraints on the factor $\Phi(\theta)$, which further restricts admissible values of $s$ in \eqref{cddsum1} to certain a subset $\{s\} \subset \mathbb{N}$. Importantly, in all known cases $\{s\}$ coincides with the set of spins of local IM \eqref{Psdef} of the given IQFT. Therefore, in the typical situation described above, the space of infinitesimal deformations of a factorizable $S$-matrix involves a
finite-dimensional part related to the deformations of solutions of the Yang-Baxter equation (in what follows we refer to those as the "principal deformations"), and also infinite-dimensional space of deformations of the CDD factor,
\begin{eqnarray}\label{cdd-deformations}
\delta{\hat S}(\theta) = \bigg(i\sum_{s\in\{s\}}
\,\delta\alpha_s\,\sinh(s\theta)\bigg)\,{\hat S}(\theta)\,.
\end{eqnarray}

On the other hand, deformations of QFT, and hence deformations of $S$-matrix, are generally generated by local fields $O$ from $ \mathcal{F}^{(0)}/\partial\mathcal{F}$. The S-matrix version of the
deformation formula \eqref{defformula} reads
\begin{eqnarray}\nonumber
&&\delta_g\,\left[_{out}\langle\,A(\theta_1')...A(\theta_M')\mid
A(\theta_1) ... A(\theta_N)\,\rangle_{in}^\text{conn}\right] =\\
\label{deltaS}&&\qquad -\sum_i\,\delta g^i\,\int\,d^2 w\,_{out}\langle\,A(\theta_1')...A(\theta_M')\mid\,O_i(w)\,\mid
A(\theta_1) ... A(\theta_N)\,\rangle_{in}^\text{conn}
\end{eqnarray}
where it is assumed for simplicity that there is a single kind
of particles which we denote $A$, and we use obvious notations for the asymptotic states\footnote{Our convention for the state normalization is $\langle A(\theta)\mid A(\theta')\rangle = (2\pi)\,\delta(\theta-\theta')$.}. The matrix elements
\begin{eqnarray}\label{ffact}
\langle\,A(\theta_1')...A(\theta_M')\mid\,\cO(w)\,\mid
A(\theta_1) ... A(\theta_N)\,\rangle^\text{conn}
\end{eqnarray}
appearing in the integrand in the r.h.s. of \eqref{deltaS} are known as the form-factors. The deformation formula \eqref{deltaS}
is written for fully connected parts of both the S-matrix element in the l.h.s. and the form-factor in the r.h.s., as the superscript "conn" indicates (of course, the same formula remains valid if one includes
all disconnected parts). In what follows we always discuss in terms of the fully connected matrix elements, and omit the superscript "conn". Also, when not important, we omit indicators {\it in/out} for the states.

In IQFT the form-factors are constrained by a system of the so called "form-factor bootstrap" (FFB) equations, which can be written in closed form, provided the factorizable ${\hat S}(\theta)$ is given. FFB equations are a system of linear functional equations, and the solutions form a vector space. The form of the FFB equations is independent of the choice of the field $O$ involved, and from this point of view $O$ can be regarded as just a tag labeling basic vectors in the vector space of solutions of FFB.
{It is generally believed  that the space of solutions of FFB equations is isomorphic to the space $\mathcal{F}$ of local fields of the IQFT.
In a number of important models this expectation was supported by counting of the solution
(see {\it e.g. }   \cite{mussardo-ff,delnic}), and for the sine-Gordon model, even explicit relations between the bases was established \cite{JMSFF,BS}.}

For generic $O_i(w)$ there is no reason for the connected matrix element in the r.h.s of \eqref{deltaS} to vanish neither at $M\neq N$, nor, if $M=N$, at $\{\theta_1',...,\theta_N'\} \neq  \{\theta_1,...,\theta_N\}$. Therefore, of course, generic $O_i \in {\hat{\mathcal{F}}}$ generates non-elastic scattering processes, and thus the corresponding deformation $\delta g^i$ breaks integrability. However, recall that the S-matrix elements always involve the energy-momentum delta functions
\begin{eqnarray}\label{deltafunct}
i\,(2\pi)^2\,\,\delta\left(\Delta P_{+}\right)\,\delta\left(\Delta P_{-}\right)
\end{eqnarray}
(which in \eqref{deltaS} emerges after the $w$-integration), where
\begin{eqnarray}\label{Ppmdef}
\Delta P_{+} = \sum_{k=1}^N \,p_{+}(\theta_k)-\sum_{l=1}^M \,p_{+}(\theta_l)\,,\qquad  \Delta P_{-} = \sum_{k=1}^N \,p_{-}(\theta_k)-\sum_{l=1}^M \,p_{-}(\theta_l)\,,
\end{eqnarray}
and $p_{\pm}(\theta) = M\,e^{\pm \theta}$. Therefore, the deformation would not generate inelastic processes
provided the form-factors in \eqref{deltaS} with $N>2$ or $M>2$ vanish on the "energy-momentum surface" supporting the delta-functions \eqref{deltafunct}. To make this property consistent with analyticity,
one then has to demand that the connected form-factors with $N>2$ or $M>2$ have the form
\begin{eqnarray}\nonumber
&&\langle A(\theta_1')...A(\theta_M')\mid O(0) \mid
A(\theta_1) ... A(\theta_N)\rangle = \\
&&\qquad\qquad\qquad\qquad  \Delta P_{+}\,G^{+}(\{\theta\}|\{\theta'\})+\Delta P_{-} \,
G^{-}(\{\theta\}|\{\theta'\})\label{Pfactor}
\end{eqnarray}
with $G^{\pm}(\{\theta\}|\{\theta'\})$ regular at the energy-momentum surface. And it is easy to see that the structure \eqref{Pfactor} is fully consistent with all FFB equations, notably with the "annihilation pole"
equation, which states that the form-factors \eqref{ffact} have poles
when one of the final rapidities $\theta_l'$ coincides with any of the initial rapidities $\theta_k$, and relates the residues of these poles to the reduced form-factors \eqref{ffact}, with the particles $A(\theta_l')$ and $A(\theta_k)$ deleted from the bra and ket states, respectively; clearly, under such reduction the structure \eqref{Pfactor} is preserved, because at $\theta_l'=\theta_k$ the associated terms in the sums in \eqref{Ppmdef} cancel out. Of course, if $O$ is a derivative of another local field, i.e. $O\in\partial\mathcal{F}$, its form-factors \eqref{ffact} vanish on the energy-momentum surface automatically, for any $N$ and $M$. And naively, one might conclude that the structure \eqref{Pfactor} suggests
$O\in\partial\mathcal{F}$. This, however, is not always the case, because
for $N=M=2$ the energy-momentum surface defined by
$\Delta P_{+}=0, \Delta P_{-}=0$ lies entirely within the locus of the annihilation poles of the form-factor
\begin{eqnarray}\label{22ffact}
_{in}\langle A(\theta_1')A(\theta_2') | O(0) | A(\theta) A(\theta_2)\rangle_{in}\,.
\end{eqnarray}
Indeed, in this case the energy-momentum conservation requires that both variables $\theta_1 -\theta_1'$ and $\theta_2-\theta_2'$ (or the same with $\theta_1\leftrightarrow \theta_2$, if the particles have equal masses) turn to zero, while one hits the annihilation pole by bringing to zero either one of these variables. Set $\theta_1'=\theta_1+\epsilon_1, \theta_2'=\theta_2+\epsilon_2$, and expand \eqref{22ffact} in double Laurent series in $\epsilon_1$ and $\epsilon_2$ (remember that the form-factor \eqref{22ffact} is a meromorphic function). It is not difficult to see from the annihilation pole residue equation of the FFB \cite{book} that the leading terms have the form
\begin{eqnarray}\label{freg}
\frac 1{i}\left(\frac{\epsilon_1}{\epsilon_2} +\frac{\epsilon_2}{\epsilon_1}\right)\,{\hat S}^{-1}(\theta_{12}){\hat S}'(\theta_{12})\,\langle
A\mid O\mid A\rangle +  {\hat f}^\text{reg}_O(\theta_{12})+O(\epsilon_1,\epsilon_2)\,,
\end{eqnarray}
with $\theta_{12}:=\theta_1-\theta_2$, and the prime denotes the derivative. The regular part ${\hat f}^\text{reg}_O$ generally does not vanish. The singular terms can be attributed to the mass operator $O$ insertions into the external legs of the $2\to 2$ scattering amplitude, as the presence of the factor $\langle A\mid O\mid A\rangle=\langle A(\theta)|O|A(\theta)\rangle$ (which in fact is $\theta$-independent constant) suggests.

From this analysis we conclude that special solutions of the FFB equations can exist, for which all form factors \eqref{ffact} with $N>2$ or $M>2$ vanish on the energy-momentum surface, but the ones with $N=2, M=2$ do not. We call those the "special form factors", and denote $X$ the operators associated with these special solutions. Clearly, adding to $X$ any total derivative of a local field generates again a special solution. Since total derivatives are totally irrelevant in the deformation formula \eqref{deltaS}, we may regard $X$ as being defined modulo total derivatives. Although a general proof is not available, it is very plausible that the space of special solutions, when factorized over the total derivatives, is isomorphic to $T\Sigma^\text{Int}$.

Every special solution $X\in {\hat{\mathcal{F}}}$ generates infinitesimal deformation of the factorizable S-matrix, by the deformation of the two-particle S-matrix (see \cite{mussardo}),
\begin{eqnarray}\label{amplitude}
{\hat S}(\theta) \ \to\ {\hat S}(\theta)\,\left[1 + \frac{i\,\delta g}{\sinh\theta}
\,\left({\hat f}^\text{reg}_X(\theta) - 2i\,{\hat S}^{-1}(\theta){\hat S}'(\theta)\,\langle A \mid X \mid A \rangle\,\cosh\theta \right)\right]
\end{eqnarray}
where ${\hat f}_X^\text{reg}(\theta)$ is the regular part in \eqref{22ffact}. On the other hand, it is natural to assume that any infinitesimal deformation of factorizable S-matrix can be generated by some special solution $X$ of the FFB equations, via \eqref{amplitude}. Thus, the deformations of the CDD factor \eqref{cdd} are generated by the operators $X_s$ (see sec. 5). Then \eqref{cdd-deformations} suggests that we must have\footnote{Note that for all deformations $X_s$ but $X_1$ the second term in \eqref{amplitude} vanishes, since $\langle A \mid X_s \mid A\rangle =0$ for all $s>1$, as is easily deduced from \eqref{nXn}. For $s=1$ the well known relation $\langle A \mid X_s \mid A\rangle =\pi M^2\, \langle \Theta \rangle$ holds.}
\begin{eqnarray}\label{fXs}
&&f_{X_s}^\text{reg}(\theta) = \kappa_s\,\sinh(\theta) \,\sinh(s\theta)\quad \text{for}\quad s>1\,,\\
&&f_{X_1}^\text{reg} = -2\pi M^2\,\langle \Theta \rangle\,\varphi(\theta) \,\cosh\theta\,\label{fX}
\end{eqnarray}
where the constants $\kappa_s$ depend on the normalization
of the currents $(T_{s+1},\Theta_{s-1})$, and $M$ is the particle's mass. This expectations will be confirmed by explicit calculations in the sine-Gordon model in the next section.

\section{Example: Sine-Gordon model}

The Sine-Gordon model,
\begin{align}\label{SG}
\mathcal{A}_\mathrm{SG}[\varphi]=\int  \left[\frac 1 {4  \pi}\, \partial _z\varphi \partial_{\bar{z}}\varphi -
\frac{\mu^2}{\sin\pi \beta^2} \cos(\beta\varphi) \right]d^2 z\
\end{align}
with real $\beta < 1$, is perhaps the best known IQFT. Here we use this example to substantiate the statements of the previous section.

The particle spectrum and factorizable S-matrix associated with the model are well known (see e.g. \cite{zz1}). Stable particles are quantum soliton $A_+$ and corresponding anti-soliton $A_-$ (which can be regarded as as basic vectors in
the space $\mathbb{C}^2$ of the charge states), and a number (which depends on the range of the coupling parameter $\beta$) of neutral "quantum breathers" $B_n$. The latter can be regarded as $A_+\,A_-$ bound states. The two-particle S-matrix of solitons and anti-solitons is an operator ${\hat S}(\theta) = S_{\epsilon_1 \epsilon_2}^{\epsilon_1' \epsilon_2'}(\theta)$ acting in the tensor product {$\mathbb{C}^2\otimes\mathbb{C}^2$} of the charge states of the scattering particles. Its explicit form can be found in Ref.\cite{zz1}.

The model \eqref{SG} can be regarded as a CFT perturbed by a relevant operator. According to arguments in \cite{Z1989}, the space of local fields $\mathcal{F}_\text{SG}$ is isomorphic to the space of fields of the associated UV CFT. It includes the exponentials $e^{ia\beta\varphi}$ with arbitrary real $a$, as well as all
their "descendants". The latter are suitably regularized fields of the form $P(\partial_\mu\varphi,\partial_\mu\partial_\nu\varphi, ...)\,e^{ia\beta\varphi}$, where $P$ are arbitrary polynomials of the first and higher derivatives of $\varphi$. The descendants of a given exponential form a bosonic Fock space $\tF_a$, where the exponential field is identified with the Fock vacuum. As in CFT, each of the spaces $\tF_a$ splits into the "level subspaces" characterized by spins and scale dimensions of the descendants. It is useful to define the spaces
\begin{eqnarray}
\mathcal{F}_a =
\oplus_{_{n=-\infty}}^{^\infty}\,\text{F}_{a+n}\,.
\end{eqnarray}
which combine all the fields with the same transformation under the symmetry $\beta\varphi\to\beta\varphi + 2\pi\mathbb{Z}$ of \eqref{SG}. The full space $\mathcal{F}_\text{SG}$ is the continuous direct sum of $\mathcal{F}_a$ with $a\in[-1/2:1/2]$.

Being IQFT, the model \eqref{SG} has an infinite number of local IM of the form \eqref{Psdef}, \eqref{Pbsdef}. In this case the spins $s$ run all positive odd values, i.e. $\{s\} = \mathbb{N}+1$. All currents $(T_{s+1},\Theta_{s-1})$, as well as the negative-spin currents, lie in the subspace $\mathcal{F}_0$. The IM $P_s$, ${\bar P}_s$ act in $\mathcal{F}_\text{SG}$ by commutators: $\forall \cO \in \mathcal{F}_\text{SG}$
\begin{eqnarray}\label{isdef}
{\bf i}_s \cO (z) := [P_s, \cO(z)]\,, \qquad {\bar{\bf i}}_s\cO(z) = [{\bar P}_s, \cO(z)]\,,
\end{eqnarray}
where the commutator may be understood as in \eqref{commutator}. In fact, the operators ${\bf i}_s, {\bar{\bf i}}_s$ act separately in each of the subspaces $\mathcal{F}_a$.

The above structure of $\mathcal{F}_\text{SG}$ was essentially proven
in Refs.\cite{HGSV,JMSFF}, starting with the lattice realization of the model. By explicit lattice construction, and then taking the continuous
limit, it was found that $\mathcal{F}_\text{SG}$ supports action of
an infinite set of fermionic "creation operators"
\begin{eqnarray}\label{betagamma}
\betab ^*_{s}\,, \  \gammab^*_{s}\quad \text{and}\quad
\bar\betab ^*_{s}, \  \bar\gammab^*_{s}\,,
\end{eqnarray}
along with the corresponding "annihilation operators" $\betab_{s}$, $\gammab_{s}$ and $\bar\betab_s$, $\bar\gammab_s$, where again $s$ runs over $2\mathbb{N}-1$. The fermionic
operators obey standard anti-commutation relations, with nonzero anti-commutators
\begin{eqnarray}
\{\betab_{s},\betab^*_{s'}\}=\delta_{s,s'},\quad \{\gammab_{s},\gammab^*_{s'}\}=\delta_{s,s'}\,,
\end{eqnarray}
and all fermions commute with ${\bf i}_s$, ${\bar{\bf{i}}}_s$ defined in \eqref{isdef}. The operators $\betab_{s}, \gammab_{s}$ annihilate the exponential fields $e^{ia\beta\varphi}$. Then, the spaces $\mathcal{F}_a$ emerge within the module $\Psi_a$ generated by the operators \eqref{isdef} and \eqref{betagamma}\footnote{Relation between this "fermionic" basis in $\mathcal{F}_a$ is complicated but can be established level by level \cite{HGSIV,NS}. The important question of explicit realization of the fermionic operators directly in the continuous theory \eqref{SG}, or even in its UV CFT, remains largely open.}. More precisely, by ascribing the "fermionic charges" $q$ to the operators \eqref{betagamma} ($q=+1$ to $\gammab^*_{s}$ and $\bar\gammab^*_{s}$, and $q=-1$ to $\betab ^*_{s}$ and $\bar\betab ^*_{s}$))  the module $\Psi_a$ can be split into the sum
of of subspaces $\Psi^{(q)}_a$ of given $q$. Then the zero charge sector $\Psi^{(0)}_a$ is isomorphic to $\mathcal{F}_a$. Form factors of the fields associated with natural basic vectors in $\Psi^{(0)}_a$ (monomials in \eqref{betagamma}) have simple compact form \cite{JMSFF}.

As was mentioned, components of conserved currents all lie in the space
$\mathcal{F}_0$. They have simple form in the fermionic basis
\begin{eqnarray}
&&T_{s+1}=C_s\,\betab^*_{s}\gammab^*_1\cdot 1\,,\qquad \Theta_{s-1}=C_s\,\betab_{s}^*\bar{\gammab}^*_1\cdot 1\,,\\
&&\Tb_{s+1}=C_s\, \bar \betab^*_{s}\bar \gammab^*_1\cdot 1\,,\qquad
\Thb_{s-1}=C_s\,\bar \betab_{s}^*{\gammab}^*_1\cdot 1\,,\nonumber
\end{eqnarray}
where $1\in \mathcal{F}_0$ is the identity field, and $C_s$ are constants whose values depend on the normalization of the currents. Furthermore, the scalar fields $X_s$ defined by \eqref{Xdef} are identified with the vectors
\begin{eqnarray}\label{Xsfermion}
X_s=C_s^2\, \betab^*_1\gammab^*_{s}\bar\betab^*_1\bar \gammab^*_{s}\cdot 1\,.
\end{eqnarray}

One way to establish \eqref{Xsfermion} is to use the following remarkable identity \cite{JMSFF}. Consider the operator
\begin{eqnarray}\label{Qdef}
{\bf Q}=\sum_{s\in \mathbb{N}-1}\left({\bf i}_{s}\gammab_{s}+{\bar{\bf i}}_{s}{\bar\gammab}_{s}\right)\,,
\end{eqnarray}
which has the fermionic charge $+1$, and squares to zero. It is possible
to show that $\bf Q$ annihilates all vectors in the $q=-1$ subspace of the $a=0$ module,
\begin{eqnarray}\label{nullvec}
{\bf Q}\Psi_{0}^{(-1)} =0\,.
\end{eqnarray}
For instance, applying \eqref{Qdef} to $\betab_s^*\gammab_1^*{\bar\gammab}_1^*\cdot 1$ leads to $\left({\bf i}_1\betab_{s}^*{\bar\gammab}_{1}^* -{\bar{\bf i}}_1\betab_{s}^*\gammab_{1}^*\right)\cdot 1 =0$, which of course is the continuity equation \eqref{Tcont}. More to the point, apply $\bf Q$ to the vector $\betab^*_{s}\bar{\betab}^*_{s}\gammab^*_1 \bar{\gammab}^*_1\gammab^*_{\sigma}\cdot 1$. Then \eqref{nullvec} yields
\begin{eqnarray}
{\bf i}_{\sigma}\betab^*_{s}\bar{\betab}^*_{s} \gammab^*_1\bar{\gammab}^*_1\cdot 1
 =
 {\bar{\bf i}}_{1}\betab^*_{s}\bar{\betab}^*_{s}\gammab^*_1\gammab^*_{\sigma}\cdot 1-\mathbf{i}_{1}\betab^*_{s}\bar{\betab}^*_{s} \bar{\gammab}^*_1\gammab^*_{\sigma}\cdot 1\,.
 \end{eqnarray}
While the l.h.s. represents the commutator of $P_\sigma$ with the field
\eqref{Xsfermion}, the r.h.s. is expressly a total derivative, in agreement with \eqref{PXcomm}. As a bonus, this calculation gives the explicit form of the fields appearing in the r.h.s of \eqref{PXcomm1},
\begin{eqnarray}
{\hat T}_{\s+1,\,s}=\betab^*_{s}\bar{\betab}^*_{s}\gammab^*_1\gammab^*_{\sigma}\cdot 1\,, \qquad {\hat\Theta}_{\s-1,\,s} = \betab^*_{s}\bar{\betab}^*_{s} \bar{\gammab}^*_1\gammab^*_{\sigma}\cdot 1\,.
\end{eqnarray}
in the sine-Gordon model.

Alternatively, the identification \eqref{Xsfermion} can be established using the explicit form of the finite-size matrix elements of these fields, which, in particular, expressly satisfy \eqref{nXn} (see \cite{HGSV}, Eq. (10.5)). Moreover, using explicit expression of the form factors in the fermionic basis (see \cite{JMSFF}), it is not difficult to verify  that the form-factors of the fields \eqref{Xsfermion} satisfy all the special properties described in in Sect.7, namely the form factors \eqref{ffact} with $N>2$ or $M>2$ vanish, while the $N=M=2$ form factors reproduce \eqref{fX}
with
$\kappa_s = C_s^2\,,$
(the last result, Eq. \eqref{fXs}, for the sinh-Gordon model was obtained independently in a recent paper \cite{Lashkevich} by a different method), and $\langle \Theta\rangle =\pi \frac{M^2}4\cot\left(\frac{\pi}{2(1-\beta^2)}\right) $.

We note here that the sine-Gordon model S-matrix has, apart from the CDD deformations generated by $X_s$, two independent "principal" deformations. One is the change in the parameter $\beta$, which is generated by the field $(\partial_\mu\varphi)^2$. The other is less obvious. It is generated by the "soliton-creating" operator $Y = \mathcal{O}_0^4 + \mathcal{O}_0^{-4}$; here and below we use the terminology and notations of Ref.\cite{LZ2001}. The two terms have soliton charges $+4$ and $-4$, respectively, therefore after deformation by this operator the theory conserves the soliton number only modulo $4$. It is possible to prove that this deformation preserves integrability, and the operator $Y$ generates the "8-vertex" deformation of the SG S-matrix which is described in Ref.\cite{Z1979}. An instructive way to show that $Y \in T\Sigma^\text{Int}|_\text{SG}$ is to recall that SG model has the symmetry with respect to the affine quantum group $U_q(\hat{SL}(2))$ symmetry \cite{Bernard-LeClair} whose generators are given in terms of non-local currents $(\mathcal{J}_\pm, \mathcal{H}_\pm)$ and $({\bar{\mathcal{J}}}_\pm, {\bar{\mathcal{H}}}_\pm)$ of fractional spins (see Eq.(4.8) of \cite{LZ2001} for definitions). Albeit non-local, the currents satisfy continuity equations \eqref{Tcont}, which allows one to derive
\begin{eqnarray}
\mathcal{J}_{\pm}(z){\bar{\mathcal{J}}}_{\pm}(z')-\mathcal{H}_{\pm}(z){\bar{\mathcal{H}}}_{\pm}(z') = \mathcal{O}^{\pm 4}_0(z') + \text{derivatives}
\end{eqnarray}
in analogy to \eqref{Xdef}, and use the arguments of Sect.6 to show that $[P_s, \mathcal{O}_0^{\pm 4}(z)] \in \partial\mathcal{F}$\footnote{By the same arguments one can show that the operators
$\mathcal{O}_{\pm 1/\beta}^0 = \exp\{\pm \frac{i}{\beta}\varphi\}$ also lie in $T\Sigma^\text{Int}|_\text{SG}$. Perturbing with these operators generates a confining interaction
between the solitons, which completely restructures the particle spectrum. There are a number
of puzzles regarding this deformation,  we do not feel ready to discuss it.}.

\section{Discussion}

Here we considered the subspace $\Sigma^\text{Int}$ of integrable QFT is the space $\Sigma$ of all QFT. Given a IQFT $\in \Sigma^\text{Int}$, we studied the content of the tangent space $T\Sigma^\text{Int}\big|_\text{IQFT}$ of infinitesimal deformations of IQFT which preserve  integrability. We found that this space contains infinitely many independent vectors $X_s$, where $s$ runs the same values that labes the local IM $P_s$ of the IQFT. The full tangent space may include finitely many additional basic vectors (see discussion at the end of sect.7). Since massive IQFT are described by their factorizable S-matrces, we observed that the local deformations $X_s$ are in correspondence with infinitesimal deformations \eqref{cdd-deformations} of the two-particle $S$-matrix by the CDD factor. The remaining admissible deformations correspond to deformations of the solutions of
Yang-Baxrer equations proper.

In the previous sections we almost completely ignored the problem of short distance behavior of QFT. We assumed that QFT constituting $\Sigma$ are equipped with a certain UV cutoff, with  cutoff distance $\epsilon$, and limited attention to the scales much greater than $\epsilon$. Note that this is exactly the space in which Wilson's RG transformations are defined, see Ref.\cite{wilson}. Of course, it is of much interest to understand which members of $\Sigma$ are "UV complete", i.e. admit meaningful limit $\epsilon \to 0$. This condition is naturally formulated in the language of RG: we are interested in RG trajectories which can be extended backward in  RG "time" (the logarithm of the characteristic length scale) indefinitely, without encountering any singularities or pathologies. Such UV complete QFT constitute but a small subspace in the whole space of QFT, $\Sigma(\infty) \subset \Sigma$ in the notations of Ref.\cite{wilson} \footnote{Of course the fact that the overwhelming majority of small deformations of UV complete QFT are not UV complete is well known since the discovery of the "Moscow Zero" in QED, and the $(T{\bar T})$-flow discussed in section 5 seems to provide an "exactly solvable" example of the UV problem generated by deformation: at ${\tilde\alpha} C <0$ the energy level $E_\alpha(R)$ develops square-root singularity at certain finite positive $R$ which may be much greater than the cutoff distance $\epsilon$. And we do not believe that just naming this singularity "the Hagedorn transition" of some sort dismisses the problem of explaining its physical nature. We hope to return to this question elsewhere \cite{SZ2}.}. Generally, characterization of $\Sigma(\infty)$ and the associated tangent space is very difficult problem even in 2D, but perhaps it can be simplified if one limits attention to IQFT. This sort of ideas was one of the main motivations for this work.

In principle, there are many ways to probe short distance behavior of a theory. Perhaps the simplest is to consider the energy spectrum of a finite size system in the geometry shown in Fig.1. Then the behavior of $E_n(R)$ at small $R$ might tell us something about short distance in a given theory. Somewhat more complicated but still feasible is to look at the two-point correlation functions of local operators, through their intermediate-state decompositions in terms of the form factors \cite{alzam}. Then the UV consistency may be probed by looking at short distance behavior of such correlation functions. {Below we mostly discuss the finite-size energies, but then make some remarks concerning the second approach.}

In massive IQFT with known factorizable S-matrix the finite-size energies can be obtained using the Thermodynamic Bethe Ansatz (TBA) method and/or its generalizations\cite{TBA,BLZ,DT}\footnote{Another approach is based on so called "Non-Linear Integral Equations" (NLIE), generalizing Destri-deVega equations \cite{ddv}. In many cases it is more powerful than TBA, but so far it lacks the universality and model independence of the TBA algorithm.}.
One can take factorizable S-matrix of any known IQFT and modify it by a CDD factor \eqref{cddsum1}, and then solve (numerically) the modified TBA equations for the ground state energy $E(R)$. In fact, the coordinates $\alpha_s$ are not the best for this sort of calculation (primarily because the series in \eqref{cddsum1} has limited domain of convergency in $\theta$); it is more convenient
to use the conventional representation
\begin{eqnarray}\label{cddpoles}
\Phi(\theta) =\prod_p^N\,\frac{B_p-i\sinh\theta}{B_p+i\sinh\theta}\,,
\end{eqnarray}
where the parameters $B_p$ are either real negative or enter in complex
conjugated pairs with negative real part (see e.g. \cite{zz1}), and the number $N$ of the factors in the product may be finite or infinite\footnote{If the number of factors is finite, such modification does not alter the high-energy asymptotic of the S-matrix, except for possibly changing the sign. However, with the infinite product the UV behavior can be substantially affected. For example, by taking the limit $\lim_{N\to \infty} \,\left[\frac{2N-i\alpha \,\sinh\theta}{2N+i\alpha\,\sinh\theta}\right]^N$ one can
obtain the exponential factor $\exp\{-i\alpha\sinh\theta\}$ appearing in the $T({\bar T})$ flow, see sect. 5. The union of the collections $\{B_p\}$ with all numbers of entries can be regarded as another coordinates in the space of CDD factors, alternative to the coordinates $\{\alpha_s\}$ in \eqref{cddsum1}.}. The union of the $N$-tuples $\{B_p, p=1,\cdots, N\}$ with all $N$ constitutes coordinate system in the space of CDD factors, alternative to the coordinates $\{\alpha_s\}$ in \eqref{cddsum1}. If ignoring the parameters associated with the principle deformations of the factorizable S-matrix, both can be regarded as coordinates in $\Sigma^\text{Int}$. Under special arrangements of the CDD poles, the TBA equations may lead to $E(R)$ which is regular at all positive $R$ and display CFT-like singularity at $R=0$, and these CDD deformations can be shown to give rise to UV complete IQFT. Famous example of this kind is provided by Al. Zamolodchikov's staircase model \cite{AlZroaming}, whose
S-matrix, from the point of view of the present paper, is the free Majorana fermions S-matrix with a simple CDD factor. Further examples can be found in \cite{DF,mart}.
But it is known for a long time that with generic choice of the CDD factor \eqref{cddpoles}, solution of the TBA equations results in $E(R)$ having singularity at finite $R$. Extensive analysis of this phenomenon was conducted by Al. Zamolodchikov in the early 90's \cite{Al-cdd}, who, by careful numerical calculations, discovered that in all cases that such singularity emerged, it happened to be the square root branching point. The singularity was later observed (without elucidating its character) in \cite{mussardo1}, and was attributed to the bosonic character of the TBA equation there. In fact, the singularity seems to be a typical feature of solutions of TBA equations, bosonic and fermionic alike. It also does not stem from any abnormality in the high-energy behavior of ${\hat S}(\theta)$, in particular, the square-root singularity appears under finite-$N$ deformations \eqref{cddpoles}, with generic choice of $B_p$. It seems suggestive to note that, at least mathematically, these singularities are of the same character as the singularities observed in $E_\alpha(R)$ in the $T{\bar T}$-flow in sect.5.  It is tempting to assume that appearance of such singularities indicates a violation of true locality, in other words that theories with such singularities lie outside $\Sigma(\infty)$.
But it is difficult to support this assumption without much better understanding of the physics behind the formation of such singularities. Some steps in this direction will be reported in \cite{SZ2}.

Alternative approach is to study the correlation functions using by known approach
based on the intermediate-state decomposition, with the use of exact form factors \cite{karowski, alzam}. For instance, the two-point function $\langle O(z) O(0) \rangle$ is represented as the sum
\begin{eqnarray}
\sum_{n=1}^\infty \,\frac{1}{n!}\,\int\,\Bigl[\prod_{j=1}^n \frac{d\theta_j}{2\pi}\Bigr]\,\big|\langle A(\theta_1) \ldots A(\theta_n)\mid O(0)\mid 0\,\rangle\big|^2\,\exp\Bigl\{-MR\,\sum_{i=1}^n \cosh\theta_i\Bigr\}
\end{eqnarray}
where $R=\sqrt{\tz\tzb}$ (again, for simplicity we assume single particle in the spectrum).
In local theories the series converges for all positive $R$ (and for all $R$ with positive real part). In principle, one can find solutions of the FFB equations for any S-matrix, and the CDD deformations lead to corresponding deformations of the form factors. For example, for the Sine-Gordon model S-matrix with CDD factors \eqref{cddpoles} the it is not too hard
to find the form factors using the technique of \cite{JMSFF}. Very little is known about convergence of the series for generic S-matrix, but it is plausible that generally, even with regular high-energy behavior, the series converges only at sufficiently large $R>R_*$, and analytic continuation shows branching point singularity at $R_*$. This of course would indicate breakdown of locality (the discontinuity across the branch cut is directly related to the commutator). The problem deserves detailed study.

In this work we only considered Lorentz invariant IQFT. But some of our main results seems to apply to more general setting. Thus, the properties established in Sect.6 for the operators $X_s$ generalize straightforwardly for the fields $X_{s,s'}(z') = \break \lim_{z\to z'} [T_{s+1}(z)\Tb_{s'+1}(z')-\Theta_{s-1}(z){\bar\Theta}_{s'-1}(z')]$, which may have non-zero spins $s-s'$, and thus generate integrable deformations breaking the Lorentz invariance.
Connections of such "effective theories" with lattice integrability seems an interesting question to explore.

\section*{Acknowledgments}

\noindent

AZ acknowledges the hospitality and support of the Simons Center of Geometry and Physics and the
Weizmann Institute of Science at various stages of this work. Discussions and insights from A.Polyakov, E.Witten, Z. Komargodski, G. Falkovich, S. Lukyanov, P. Vieira and S. Dubovski
were highly appreciated. Research of AZ is supported by DOE grant SC0010008.

FS and AZ are grateful to the Perimeter Institute for hospitality in June/July 2016, when
the manuscript was essentially completed. We thank A.Cavagli\'a, S.Negro, I.Szecsenyi, and R.Tateo for sharing their results prior to publication.

\end{document}